\documentclass[sigconf,nonacm]{acmart}

\AtBeginDocument{%
  \providecommand\BibTeX{{%
    \normalfont B\kern-0.5em{\scshape i\kern-0.25em b}\kern-0.8em\TeX}}}

\usepackage{hyperref}

\begin{document}

\title{pymovements: A Python Package for~Eye~Movement~Data~Processing}

\author{Daniel G. Krakowczyk}
\email{daniel.krakowczyk@uni-potsdam.de}
\orcid{0009-0009-5100-0733}
\affiliation{%
  \institution{University of Potsdam}
  \city{Potsdam}
  \country{Germany}
}

\author{David R. Reich}
\email{david.reich@uni-potsdam.de}
\orcid{0000-0002-3524-3788}
\affiliation{%
  \institution{University of Potsdam}
  \city{Potsdam}
  \country{Germany}
}

\author{Jakob Chwastek}
\email{jakob.chwastek@uni-potsdam.de}
\orcid{0000-0001-7092-6245}
\affiliation{%
  \institution{University of Potsdam}
  \city{Potsdam}
  \country{Germany}
}

\author{Deborah N. Jakobi}
\email{deborahnoemie.jakobi@uzh.ch}
\orcid{0000-0002-9719-6673}
\affiliation{%
  \institution{University of Zurich}
  \city{Zurich}
  \country{Switzerland}
}

\author{Paul Prasse}
\email{paul.prasse@uni-potsdam.de}
\orcid{0000-0003-1842-3645}
\affiliation{%
  \institution{University of Potsdam}
  \city{Potsdam}
  \country{Germany}
}

\author{Assunta Süss}
\email{assunta.suess@uni-potsdam.de}
\orcid{0000-0002-9797-0278}
\affiliation{%
  \institution{University of Potsdam}
  \city{Potsdam}
  \country{Germany}
}

\author{Oleksii Turuta}
\email{oleksii.turuta@nure.ua}
\orcid{0000-0002-0970-8617}
\affiliation{%
  %\institution{Kharkiv NURE}
  \institution{Kharkiv National University of Radio Electronics}
  \city{Kharkiv}
  \country{Ukraine}
}

\author{Paweł Kasprowski}
\email{pawel.kasprowski@polsl.pl}
\orcid{0000-0002-2090-335X}
\affiliation{%
  \institution{Silesian University of Technology}
  \city{Gliwice}
  \country{Poland}
}

\author{Lena A. Jäger}
\email{jaeger@cl.uzh.ch}
\orcid{0000-0001-9018-9713}
\affiliation{%
  \institution{University of Zurich}
  \city{Zurich}
  \country{Switzerland}\\
  \institution{University of Potsdam}
  \city{Potsdam}
  \country{Germany}
}

\renewcommand{\shortauthors}{Krakowczyk et al.}

\begin{abstract}
    We introduce \emph{pymovements}: a Python package for analyzing eye-tracking data that follows best practices in software development, including rigorous testing and adherence to coding standards. The package provides functionality for key processes along the entire preprocessing pipeline. This includes parsing of eye tracker data files, transforming positional data into velocity data, detecting gaze events like saccades and fixations, computing event properties like saccade amplitude and fixational dispersion and visualizing data and results with several types of plotting methods. Moreover, \emph{pymovements} also provides an easily accessible interface for downloading and processing publicly available datasets. Additionally, we emphasize how rigorous testing in scientific software packages  is critical to the reproducibility and transparency of research, enabling other researchers to verify and build upon previous findings.
\end{abstract}

\begin{CCSXML}
<ccs2012>
   <concept>
       <concept_id>10010405.10010455.10010459</concept_id>
       <concept_desc>Applied computing~Psychology</concept_desc>
       <concept_significance>500</concept_significance>
       </concept>
 </ccs2012>
\end{CCSXML}

\ccsdesc[500]{Applied computing~Psychology}

\keywords{eye movements, preprocessing, event detection, software packages, scientific computing}

\maketitle

\section{Introduction}
Eye movements are a valuable tool in understanding viewers' or readers' cognitive processes, state of mind, or other characteristics as they are considered ``a window to the mind and brain'' \cite{vanGompel2007window}. They are widely used across disciplines, including  cognitive psychology, linguistics, neuroscience, and human-computer interaction. Hence, eye tracking methodology in general and the processing and analysis of eye tracking data in particular is an interdisciplinary field of research that is essential for the above mentioned disciplines. 

Due to its ease of use and flexibility, Python has become a popular choice among researchers for processing and analyzing eye movement data. 
In this paper, we introduce a new Python package called \emph{pymovements} that provides a set of tools for the processing and analysis of eye movement data.

Data preprocessing functionality of the \emph{pymovements} package includes conversion between positional and velocity gaze data and detection of gaze events like fixations and saccades by using dis\-persion-based and velocity-based methods. 
Users can also plot eye movement data using various visualization functions, including heatmaps and trace plots. 

The \emph{pymovements} package also contains a small library of publicly available datasets that can be downloaded using an accessible interface. 
The package is designed to be extensible, with additional datasets expected to be added in the near future. 
The source code is available online at \href{https://github.com/aeye-lab/pymovements}{https://github.com/aeye-lab/pymovements} and is licensed under the MIT License.

\section{Related Work}
There exist numerous other Python packages and tools that can be used to process or analyze eye movement data~\cite{cili,GazeParser,Perception-Engineers-Toolbox,PyTrack-NTU,sideeye}. %PyEEGLab
However, all these packages lack thorough testing and a comprehensive documentation which are prerequisites for trustworthy results in scientific research.
In contrast, \emph{pymovements} applies a rigorous testing scheme following best practices in software engineering including an extensive documentation with assisting tutorials. 

Moreover, none of these packages are specifically optimized for performance.
\emph{pymovements} on the other hand leverages features from the \emph{Polars} dataframe package, which is programmed in Rust and one of the fastest dataframe software available~\cite{polars-benchmark}.

Finally, the initiative to provide and maintain a library of publicly available datasets with eye movement data is a distinguishing feature of \emph{pymovements}.

\section{Package Design}

\emph{pymovements} is structured around a central module, the dataset module, which provides a unified interface for working with eye movement datasets.
This module makes it easy to load eye movement data and save relevant results for a given eye movement study.
The dataset class provides access to the key functionality needed for preprocessing:
from transforming positional signals into velocity signals, over detecting gaze events such as (micro-)saccades and fixations,
up to analyzing specific properties for detected events, such as saccade amplitude or fixational dispersion.

The dataset class is very flexible to allow for a seamless integration of own local datasets. 
Each dataset object holds recording session parameters, gaze dataframes for raw eye gaze signals and event dataframes, which contain the detected events within the eye gaze signals.
These dataframes are organized in a consistent manner, making it easy to work with multiple datasets and compare results across studies.

\emph{pymovements} also provides a simple interface to access publicly available datasets.
Except for an example toy dataset, all these public datasets are not part of the \emph{pymovements} package,
but are definitions on where to find dataset resources online with additional details on how to load the input data to be consistent across several datasets.
This implementation is aimed to be simple enough to facilitate the inclusion of additional datasets, even to researchers without any Python programming experience.
The \emph{pymovements} package then automatically takes care of downloading the particular dataset.
%Correct mapping of dataframe columns and data types is supported out of the box.

\section{Documentation}

\emph{pymovements} comes with a comprehensive online documentation that explains each function and module in detail along with examples of typical usage.
In addition to this, we provide detailed and comprehensive step-by-step tutorials that guide users through the entire data processing pipeline  enabling them to use \emph{pymovements} to its full potential.
These tutorials cover key use cases in eye movement data processing and are accompanied by a small example toy dataset.
The documentation is automatically generated from the source code and uploaded to the web hoster with each new update, such that it always reflects the most recent state of the package.
For each update all examples and tutorials are verified to run successfully and to produce the expected results.

\section{Testing}

Testing is one of the most important software development processes to maintain high quality and consistency of the software.
\emph{pymovements} has a robust testing pipeline to ensure that additions to the codebase are thoroughly tested and reviewed before being accepted.
Every proposed change must have at least one approving review to be incorporated and must pass all existing tests and checks of the testing pipeline.
We enforce a 100\% test coverage and a consistent coding style by obligatory linting and type checking.

\section{Integration with Existing Software}

The \emph{pymovements} package supports native integration with other Python libraries, such as \emph{Polars}, \emph{Pandas}, \emph{NumPy}, and \emph{Matplotlib}.
All of the obtained data during processing can be easily exported into commonly used file formats such as CSV and Feather files.
Furthermore, \emph{pymovements} is compatible with the R programming language through the \emph{reticulate} package.
This means that researchers can use \emph{pymovements'} functionality within the R environment, 
allowing for seamless integration with R's extensive library of statistical tools and visualization packages.

\section{Future Work}

\emph{pymovements} is an evolving package that is continuously being improved and expanded to meet the needs of researchers.
We plan to add functionality to allow researchers to analyze reading behavior and expand the package's capabilities by including visual stimuli such as image and video data into the analysis pipeline.

To increase the versatility of the package, we plan to add more event detection algorithms and public dataset definitions.
Additionally, we are working on adding parsing support for more eye trackers, enabling researchers to process and analyze data from a wider range of eye-tracking systems.

\section{Broader Impact}
One of the key benefits of \emph{pymovements} is its ability to unify the work across different research groups working with eye movement preprocessing pipelines.
By providing a standardized rigorously tested interface, \emph{pymovements} helps to ensure that analyses can be easily reproduced and verified by other researchers, increasing the transparency and reliability of eye movement research.
%This also helps to facilitate cross-disciplinary collaboration and promotes the sharing of research findings.
Due to its permissive open-source license researchers can contribute their own methods and datasets, increasing both the visibility of their own work and the feature set of the software.
\emph{pymovements} has thus the potential to have a broad impact on the field of eye movement research, facilitating greater collaboration and reproducibility across different research groups.

\begin{acks}
This work was partially funded by the German Federal Ministry of Education and Research (grant 01IS20043)
and is based upon work from COST Action MultiplEYE, CA21131, supported by COST (European Cooperation in Science and Technology).
\end{acks}

\bibliographystyle{ACM-Reference-Format}
\bibliography{bibliography}

\end{document}